\documentstyle [12pt,epic,eepic]{article}
\input feynman
\begin{document}
\hfill UTHEP--94--1001\vskip.01truein
\hfill{October 1994}\vskip1truein
\def\fnote#1#2{
\begingroup\def\thefootnote{#1}\footnote{#2}\addtocounter{footnote}{-1}
\endgroup}
\def\dslash{\not{\hbox{\kern-2pt $\partial$}}}
\def\eslash{\not{\hbox{\kern-2pt $\epsilon$}}}
\def\Dslash{\not{\hbox{\kern-4pt $D$}}}
\def\Aslash{\not{\hbox{\kern-4pt $A$}}}
\def\Qslash{\not{\hbox{\kern-4pt $Q$}}}
\def\pslash{\not{\hbox{\kern-2.3pt $p$}}}
\def\kslash{\not{\hbox{\kern-2.3pt $k$}}}
\def\qslash{\not{\hbox{\kern-2.3pt $q$}}}
\def\np#1{{\sl Nucl.~Phys.~\bf B#1}}\def\pl#1{{\sl Phys.~Lett.~\bf #1B}}
\def\pr#1{{\sl Phys.~Rev.~\bf D#1}}\def\prl#1{{\sl Phys.~Rev.~Lett.~\bf
#1}}
\def\cpc#1{{\sl Comp.~Phys.~Comm.~\bf #1}}
\def\anp#1{{\sl Ann.~Phys.~(NY) \bf #1}}\def\etal{{\it et al.}}
\def\half{{\textstyle{1\over2}}}
\def\be{\begin{equation}}\def\ee{\end{equation}}
\def\ba{\begin{array}}\def\ea{\end{array}}
\def\tr{{\rm tr}}
\centerline{\Large Electroweak corrections to the toponium decay width}
\vskip.8truein
\centerline{\sc George Siopsis\fnote{\ast}{e-mail: {\tt
siopsis@panacea.phys.utk.edu}}}
\vskip.5truein
\centerline{\it Department of Physics and Astronomy}
\centerline{\it The University of Tennessee, Knoxville, TN 37996--1200}
\centerline{\it U. S. A.}\vskip.05truein\baselineskip=21pt\vskip.6truein
\centerline{\bf ABSTRACT}\vskip.2truein\par\let\sstl=\scriptscriptstyle
We discuss one-loop electroweak corrections to the decay width of toponium.
We calculate the energy-level shifts by expanding around the solution of the
Bethe-Salpeter equation in the instantaneous approximation, in analogy with
the positronium case. We show that
first-order electroweak effects are suppressed by at least four powers of the
strong coupling constant, and are therefore
negligible compared with QCD corrections. The calculation is manifestly
gauge invariant and takes into account the contributions to the decay rate
due to both Coulomb enhancement and phase space reduction effects.
\par\renewcommand\thepage{}
\vfill\eject\parskip.1truein
\pagenumbering{arabic}\par


The threshold production of a t\=t pair at future $e^+e^-$ colliders has
attracted a lot of attention~\cite{rf1}, because the measurement of its
cross-section will permit a simultaneous determination of the mass of the
top quark $m_t$, its decay width $\Gamma_t$, and the strong coupling
constant $\alpha_s (m_t)$~\cite{rf2}. The threshold regime is dominated by
t\=t resonances which enhance the cross-section. These bound states differ
from those of the other heavy quarks (b and c), because $m_t$ is so large
that the dominant decay mode of the top quark is the weak decay $t\to W^+b$.
Thus, the decay width of a top quark is
\be \Gamma_t \sim {G_F\over \sqrt 2} {m_t^3 \over 8\pi}\;. \ee
For $m_t \simeq 174~GeV$, we obtain a decay rate of the order of 1 GeV,
which exceeds $\Lambda_{QCD}$, and permits the use of perturbation theory,
because the t\=t pair will decay before it has time to hadronize. Moreover,
the quarks are produced with little kinetic energy, so that non-relativistic
techniques are applicable. Thus, the toponium energy levels are well
approximated by the Bohr levels of a QCD Coulomb-like effective potential.
To calculate the cross-section, one can determine the appropriate
Green function which obeys a non-relativistic Lippmann-Schwinger equation
with the QCD effective potential~\cite{rf2}.

An important parameter entering the Lippmann-Schwinger equation is the decay
width of toponium, $\Gamma_{t\bar t}$. In general, it can be a function of
the momentum of the top quark, and its precise form is crucial for an
accurate theoretical prediction of the cross-section
$\sigma (e^+e^-\to t\bar t)$.
To zeroth order,
\be\Gamma_{t\bar t} = 2\Gamma_t \;.\label{eq1}\ee
This relation is modified by three effects:
{\em (a)} time dilatation (the top quark lives longer in the center-of-mass
frame),
{\em (b)} phase space reduction (due to the binding energy of the quarks),
and {\em (c)} Coulomb enhancement (the b and \=b quarks should be described
by Coulomb wave functions rather than plane waves)~\cite{rf1}.
The first two effects can be easily accounted for and they lead to a
significant modification of Eq.~(\ref{eq1}). The third effect is much harder
to take into account. As was shown by Kummer and M\"odritsch, it cancels the
second effect to $o\, (\alpha_s^2)$, thus leaving time
dilatation as the only effect modifying Eq.~(\ref{eq1}), in analogy with the
case of muonium in nuclei~\cite{rf3}. We therefore obtain
\be \Gamma_{t\bar t} = 2\Gamma_t \left( 1-\langle {\bf p}^2/ m_t^2 \rangle
\right)^{1/2} \;,\label{eq2}\ee
which is only a small modification of Eq.~(\ref{eq1}). To arrive at this
result, M\"odritsch and Kummer~\cite{rf4} calculated one-loop graphs with
insertions of bound-state wave functions, as prescribed by the rigorous
Bethe-Salpeter formalism~\cite{rf5}.
The cancellation of the two effects (Coulomb enhancement and phase space
reduction) was due to non-trivial cancellations between Green functions,
which went well beyond the cancellation of gauge-dependent pieces implied by
Ward identities.

Our purpose here is to extend the results of ref.~\cite{rf4} to higher
orders in the strong coupling constant. We shall show that electroweak
corrections enter at an order higher than $o\, (\alpha_s^5)$, and therefore
we need second-order bound state perturbation theory in order to calculate
them. They are due to corrections to the
vertex function, which do not contain a contribution to the magnetic moment
of the top quark. This is in contrast to the positronium case, where
magnetic moment effects are $o\, (\alpha^5)$. The discrepancy arises from
the fact that $W$ only couples to a left-handed current.

%
\begin{figure}[h]
\unitlength=1.0pt
\begin{picture}(565,130)(90,700)
\thicklines\filltype{shade}
\put(320.00,765.00){\circle*{50}}
\put(420.00,765.00){\circle{50}}
\put(165.00,765.00){\circle{50}}
\put(165.00,765.00){\makebox(0,0)[cc]{\Large $\chi_P$}}
\put(110.00,820.00){\makebox(0,0)[cc]{${P\over 2} +p$}}
\put(110.00,710.00){\makebox(0,0)[cc]{${P\over 2} -p$}}
\put(260,765){\makebox(0,0)[cc]{\Large $+$}}
\put(265.00,820.00){\makebox(0,0)[cc]{${P\over 2} +p$}}
\put(265.00,710.00){\makebox(0,0)[cc]{${P\over 2} -p$}}
\put(320.00,765.00){\makebox(0,0)[cc]{\Large \bf V}}
\put(377.00,800.00){\makebox(0,0)[cc]{${P\over 2} +p'$}}
\put(377.00,730.00){\makebox(0,0)[cc]{${P\over 2} -p'$}}
\put(420.00,765.00){\makebox(0,0)[cc]{\Large $\chi_P$}}
\put(530,765){\makebox(0,0)[cc]{\large $=\quad 0$}}
\put(337.7,747.3){\line( 1, 0){ 30.5}}
\put(402.3,747.3){\vector( -1, 0){ 34.64}}
\put(337.7,782.7){\line( 1, 0){ 30.5}}
\put(402.3,782.7){\vector( -1, 0){ 34.64}}
\put(147.3,747.3){\vector(-2,-1){ 40}}
\put(129.4,738.4){\makebox(0,0)[cc]{$|$}}
\put(147.3,782.7){\vector(-2, 1){ 40}}
\put(129.4,791.6){\makebox(0,0)[cc]{$|$}}
\put(302.3,747.3){\vector(-2,-1){ 40}}
\put(284.4,738.4){\makebox(0,0)[cc]{$|$}}
\put(302.3,782.7){\vector(-2,1){ 40}}
\put(284.4,791.6){\makebox(0,0)[cc]{$|$}}
\dashline[+30]{3}(190,765)(230,765)
\put(210,765){\vector( -1, 0){ 0}}
\put(210,775){\makebox(0,0)[cc]{$P$}}
\dashline[+30]{3}(445,765)(485,765)
\put(465,765){\vector( -1, 0){ 0}}
\put(465,775){\makebox(0,0)[cc]{$P$}}
\end{picture}
\caption{\em The homogeneous Bethe-Salpeter equation.}
\label{fig1}
\end{figure}
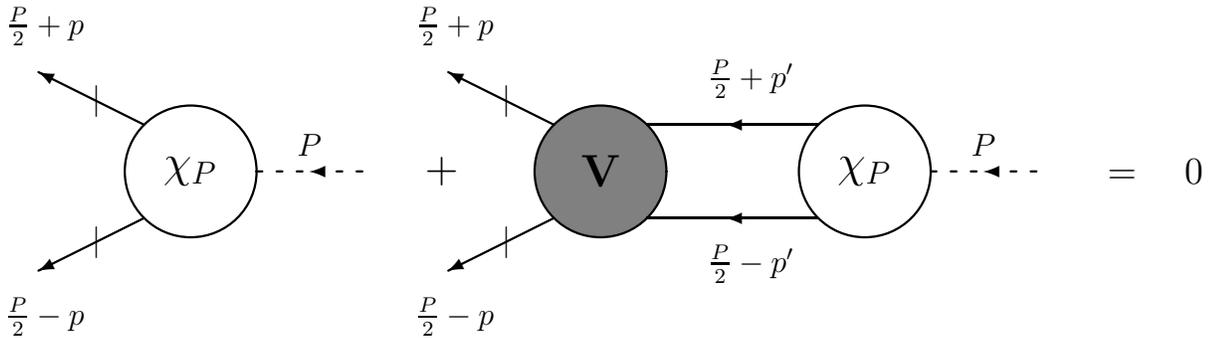
We start with a brief review of the Bethe-Salpeter formalism in order to fix
the notation. We wish to calculate the energy levels of t\=t bound states. They
are poles in the four-point amplitude describing t\=t scattering. We shall
only consider scattering in the $t$-channel, because the annihilation diagrams
are suppressed \cite{rf2}. We shall also neglect photon and $Z^0$ exchanges
because they are small effects compared to a gluon exchange.
A Higgs exchange will also
be neglected, but depending on the mass of the Higgs boson, it can have an
appreciable effect.

A bound-state wavefunction $\chi_P (p)$ satisfies the homogeneous
Bethe-Salpeter equation~\cite{rf6}
\be
\Pi^{(1)} (p_+) \Pi^{(2)} (p_-)\chi_P(p) + \int {d^4 p' \over (2\pi )^4}
V(p,p';P) \chi_P (p') =0 \;,
\label{eq3}\ee
where $\Pi(p)$ is the complete inverse fermion propagator,
\be
\Pi (p) = \pslash -M_t -\Sigma (p) \;,
\label{eq4}\ee
and we have defined momenta
\be p_\pm = {P\over 2} \pm p \;.\label{eq4a}\ee
Eq.~(\ref{eq3}) is represented graphically in Fig.~\ref{fig1}.
$V(p,p';P)$ is a potential function which consists of the two-fermion
irreducible graphs. For our purposes, the mass is complex,
\be M_t= m_t + i\Gamma_t \;. \label{eq5} \ee
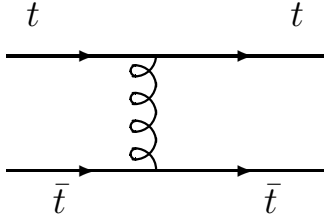
\begin{figure}
\begin{center}
\vskip 0.5in
\begin{picture}(1600,-1000)
\thicklines\filltype{shade}
\drawline\gluon[\S\REG](-400,1400)[4]
\advance\gluonfrontx by 400
\advance\gluonbackx by 400
\drawline\fermion[\E\REG](\gluonfrontx,\gluonfronty)[6000]
\drawarrow[\E\ATBASE](\pmidx,\pmidy)
\drawline\fermion[\W\REG](\gluonfrontx,\gluonfronty)[6000]
\drawarrow[\E\ATBASE](\pmidx,\pmidy)
\drawline\fermion[\E\REG](\gluonbackx,\gluonbacky)[6000]
\drawarrow[\E\ATBASE](\pmidx,\pmidy)
\drawline\fermion[\W\REG](\gluonbackx,\gluonbacky)[6000]
\drawarrow[\E\ATBASE](\pmidx,\pmidy)
\put(-5000,3000){\makebox(0,0)[cc]{\large $t$}}
\put(5000,3000){\makebox(0,0)[cc]{\large $t$}}
\put(-4000,-4000){\makebox(0,0)[cc]{\large $\bar t$}}
\put(4000,-4000){\makebox(0,0)[cc]{\large $\bar t$}}
\end{picture}
\end{center}
\vskip 0.5in
\caption{\em The zeroth-order contribution to the QCD potential.}
\label{fig2}
\end{figure}
To lowest order in $\alpha_s$ and neglecting electroweak interactions,
the potential is (Fig.~\ref{fig2})
\be V_0 (p,p';P) = C_F \, 4\pi\alpha_s \gamma_\mu^{(1)} G^{\mu\nu}
(p-p') \gamma_\nu^{(2)} \;, \label{eq6} \ee
where $C_F=4/3$ is the Casimir operator, and
$G^{\mu\nu} (k)$ is the lowest-order gluon propagator. In the Feynman
gauge (omitting group theory factors),
\be G^{\mu\nu} (k) = {\eta^{\mu\nu} \over k^2 +i\epsilon } \;, \label{eq7} \ee
and the potential $V_0 (p,p';P)$ is independent of $P$.
At threshold, the quarks move with non-relativistic velocities and the
Bethe-Salpeter equation can be approximated by the non-relativistic
Schr\"odinger equation in momentum space, and then solved. To this end,
we shall work in the total rest frame in which the overall momentum is
$P^\mu = (E, \vec 0)$. In the instantaneous approximation, the potential
becomes
\be V_0^{inst}(p,p';P) = C_F \, 4\pi\alpha_s \gamma_0^{(1)} {1\over
(\vec p-{\vec p\, }')^2} \gamma_0^{(2)} \;. \label{eq6a} \ee
If we integrate over $p^0$, we can write the Bethe-Salpeter equation
(\ref{eq3}) in terms of the wavefunction $\Phi (\vec p) = \int {dp^0 \over
2\pi} \phi (p)$ as
\be (H^{(1)} + H^{(2)} - E) \Phi (\vec p) = \left( \Lambda_+^{(1)}
\Lambda_+^{(2)} - \Lambda_-^{(1)} \Lambda_-^{(2)} \right) C_F \,
4\pi\alpha_s \int {d^3 p\over (2\pi)^3} {1\over (\vec p-{\vec p\, }')^2}
\Phi ({\vec p\, }') \;, \label{eq6b}\ee
where $H$ is the Dirac Hamiltonian and $\Lambda_+$ ($\Lambda_-$) is the
projection operator onto positive (negative) energy states.
In the non-relativistic limit, this reduces to the Schr\"odinger equation
in momentum space
\be \left( {{\vec p\, }^2 \over M_t} +2M_t- E \right) \Phi (\vec p) = C_F \,
4\pi\alpha_s \int {d^3 p\over (2\pi)^3} {1\over (\vec p-{\vec p\, }')^2}
\Phi ({\vec p\, }') \;. \label{eq6c}\ee
Thus, we obtain the energy levels
\be E_n = 2M_t - {M_t C_F^2\alpha_s^2 \over 4n^2} + o(\alpha_s^4 ) \;,
\label{eq8} \ee
which are the Bohr levels of the Coulomb-like QCD potential (\ref{eq6a}).
Therefore, the first-order QCD correction to the decay rate of toponium is
\be \Gamma_{t\bar t} = 2\Gamma_t \left( 1- {C_F^2\alpha_s^2 \over 8n^2}\right)
\;, \label{eq9} \ee
in agreement with Eq.~(\ref{eq2}) (see \cite{rf4}).
The spherically symmetric $S=0$ states are given by
\be \Phi_n (\vec p) = (M_tC_F \alpha_s)^{-3/2} {{\cal L}_n (n^2y) \over
(1+n^2y)^{n+1}}\;\;, \;\;
y={4{\vec p \, }^2 \over M_t^2 C_F^2 \alpha_s^2} \,, \label{eq9a}\ee
where ${\cal L}_n$ is a polynomial of order $n-1$ related to the Laguerre
polynomials. For $n=1$, we have ${\cal L}_1= 16\sqrt{2\pi}$.

Higher-order
corrections can be systematically introduced by perturbing around the
solution to the Schr\"odinger equation (\ref{eq6c}). The potential to be
treated perturbatively is $V-V_0^{inst}$. There is also a contribution from
the disconnected diagrams which are due to the self-energy terms in the
fermion propagators (Eq.~(\ref{eq4})), but they can be absorbed in the
potential if we make use of the Schr\"odinger equation. Thus, according to
the Bethe-Salpeter formalism \cite{rf6}, the first-order energy level
shift is
\be \Delta E_n = \langle \Phi_n |\, D_P (p)\, (H^{(1)} +H^{(2)}-E_n)\,
(V-V_0^{inst})\, (H^{(1)} +H^{(2)}-E_n)\, D_P (p)\, | \Phi_n \rangle \;,
\label{eq6d}\ee
where the inner product involves an integral over the four-momentum. $H$ is
the Dirac Hamiltonian, and
$D_P$ is the product of two free fermion propagators
({\em cf.}~Eq.~(\ref{eq3})), which can be expressed
in terms of the projection operators $\Lambda_\pm$ as
\be D_E (p) = \sum_{\pm \pm} {\Lambda_\pm^{(1)}\Lambda_\pm^{(2)} \over
[E/2 + p^0 \pm (E_p -i\epsilon )][E/2 - p^0 \pm (E_p -i\epsilon )]} \;,
\label{eq6e}\ee
where $E_p = \sqrt{{\vec p \, }^2 +M_t^2}$ is the energy of the quark on the
mass shell.

To lowest order, the
potential is $V_0-V_0^{inst}$ (Eqs.~(\ref{eq6}) and (\ref{eq6a})). This is
analogous to the positronium case, and produces an $o(\alpha_s^4)$ shift in
the energy levels.
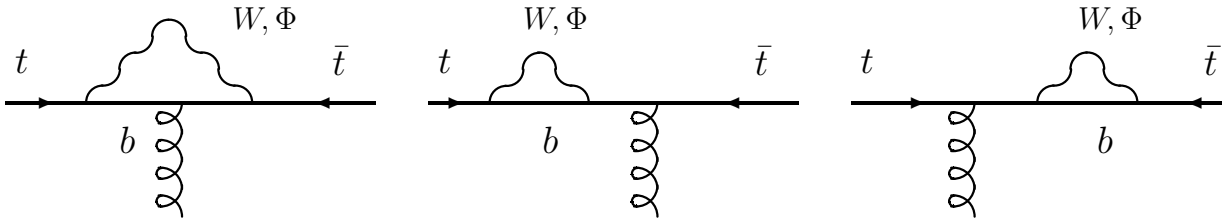
\begin{figure}
\begin{center}
\vskip 1in
\begin{picture}(1600,-1000)
\thicklines
\drawline\gluon[\S\REG](-16000,2400)[4]
\advance\gluonfrontx by 400
\drawline\fermion[\E\REG](\gluonfrontx,\gluonfronty)[3000]
\drawline\fermion[\E\REG](\fermionbackx,\fermionbacky)[4000]
\drawarrow[\W\ATBASE](\pmidx,\pmidy)
\drawline\fermion[\W\REG](\gluonfrontx,\gluonfronty)[4000]
\drawline\photon[\NE\REG](\fermionbackx,\fermionbacky)[5]
\drawline\photon[\SE\REG](\photonbackx,\photonbacky)[5]
\drawline\fermion[\W\REG](\fermionbackx,\fermionbacky)[3000]
\drawarrow[\E\ATBASE](\pmidx,\pmidy)
\put(-22000,4000){\makebox(0,0)[cc]{\large $t$}}
\put(-10000,4000){\makebox(0,0)[cc]{\large $\bar t$}}
\put(-13000,5500){\makebox(0,0)[cc]{ $W,\Phi$}}
\put(-18000,1000){\makebox(0,0)[cc]{\large $b$}}
\drawline\gluon[\S\REG](2000,2400)[4]
\advance\gluonfrontx by 400
\drawline\fermion[\E\REG](\gluonfrontx,\gluonfronty)[5000]
\drawarrow[\W\ATBASE](\pmidx,\pmidy)
\drawline\fermion[\W\REG](\gluonfrontx,\gluonfronty)[3000]
\drawline\photon[\NW\FLIPPED](\fermionbackx,\fermionbacky)[3]
\drawline\fermion[\W\REG](\fermionbackx,\fermionbacky)[4000]
\drawline\photon[\SW\FLIPPED](\photonbackx,\photonbacky)[3]
\drawline\fermion[\W\REG](\fermionbackx,\fermionbacky)[2000]
\drawarrow[\E\ATBASE](\pmidx,\pmidy)
\put(-6000,4000){\makebox(0,0)[cc]{\large $t$}}
\put(6000,4000){\makebox(0,0)[cc]{\large $\bar t$}}
\put(-2000,5500){\makebox(0,0)[cc]{ $W,\Phi$}}
\put(-2000,1000){\makebox(0,0)[cc]{\large $b$}}
\drawline\gluon[\S\REG](14000,2400)[4]
\advance\gluonfrontx by 400
\drawline\fermion[\E\REG](\gluonfrontx,\gluonfronty)[2000]
\drawline\photon[\NE\REG](\fermionbackx,\fermionbacky)[3]
\drawline\fermion[\E\REG](\fermionbackx,\fermionbacky)[5000]
\drawline\photon[\SE\REG](\photonbackx,\photonbacky)[3]
\drawline\fermion[\E\REG](\fermionbackx,\fermionbacky)[2000]
\drawarrow[\W\ATBASE](\pmidx,\pmidy)
\drawline\fermion[\W\REG](\gluonfrontx,\gluonfronty)[5000]
\drawarrow[\E\ATBASE](\pmidx,\pmidy)
\put(19000,5500){\makebox(0,0)[cc]{ $W,\Phi$}}
\put(10000,4000){\makebox(0,0)[cc]{\large $t$}}
\put(23000,4000){\makebox(0,0)[cc]{\large $\bar t$}}
\put(19000,1000){\makebox(0,0)[cc]{\large $b$}}
\end{picture}
\end{center}
\vskip 0.25in
\caption{\em The first-order electroweak correction to the three-point
vertex function contributing to the t\=t decay rate.}
\label{fig3}
\end{figure}
The first-order electroweak correction is
\be V_1 (p,p';P) = 4\pi C_F \alpha_s\alpha_W\bigg(
\Lambda_\mu^{(1)} (p_+,p_+') G^{\mu\nu}
(p-p') \gamma_\nu^{(2)} + \gamma_\mu^{(1)}G^{\mu\nu}
(p-p')\Lambda_\nu^{(2)}(p_-,p_-')\bigg) \;, \label{eq12}\ee
where $p_\pm = P/2 \pm p$, $p_\pm' = P/2 \pm p'$, and we have
made explicit the electroweak coupling constant $\alpha_W \sim G_FM_W^2$,
where $G_F$ is the Fermi constant and $M_W$ is the mass of the $W$ boson.
The vertex function $\Lambda_\mu (p,p')$ consists of the diagrams shown in
fig.~\ref{fig3}. It is guaranteed to give a gauge invariant contribution by
the Ward identity satisfied by the one-particle irreducible function,
\be (p-p')^\mu \Gamma_\mu (p,p') = \Pi (p) - \Pi (p') \;. \ee
Since we are only interested in first-order corrections, we may replace $M_t$
by its real part $m_t$.
The contribution of $V_1$ to the energy level shift (Eq.~(\ref{eq6d})) can then
be written as
$$\Delta E_n^W = {C_F^2 \alpha_s^2 \alpha_W \over 16m_t} \int {d^4 p \over
(2\pi )^4} {d^4 p' \over (2\pi )^4} {\eta^{\mu\nu} \over (p-p')^2 + i\epsilon}
\; {{\cal L}_n (n^2y) \over (1+n^2y)^n}\;
{{\cal L}_n (n^2y') \over (1+n^2y')^n} \hskip 1in$$
\be \hskip 1in\times \left\langle D_P (p') \bigg(
\Lambda_\mu^{(1)} (p_+,p_+') \gamma_\nu^{(2)} + \gamma_\mu^{(1)}
\Lambda_\nu^{(2)}(p_-,p_-')\bigg) D_P (p) \right\rangle \;, \label{nn1}
\ee
where $y=4{\vec p \, }^2 / m_t^2 C_F^2 \alpha_s^2$ and
$y'=4{\vec p \, }'^2 / m_t^2 C_F^2 \alpha_s^2$. A simple scaling argument shows
that the lowest-order contribution to the integral comes from the small
three-momentum region. Momentum insertions contribute additional powers of
$\alpha_s$. At low momentum transfer, the three-point vertex $\Lambda_\mu$ may
be written in general as
\be \Lambda_\mu (p,p') = k^2 {\cal F}_1 (k^2) + \sigma_{\mu\nu} k^\nu
{\cal F}_2 (k^2) \;, \ee
where $k=p-p'$, and $\sigma_{\mu\nu}={i\over 2}[\gamma_\mu \,,\,\gamma_\nu ]$.
The form factors ${\cal F}_1$ and ${\cal F}_2$ are regular
as $k^2 \to 0$. In the positronium case, ${\cal F}_2$ gives an $o(\alpha^5)$
contribution to the energy level shift, and is due to the magnetic moment
interaction. In our case, we need to multiply the gamma matrices by the
projection operator
${1\over 2} (1-\gamma_5)$, due to parity violation of weak interactions.
A straightforward explicit calculation shows that the form factor
${\cal F}_2 (k^2)$ vanishes to lowest order in $k^2$.
It follows that the three-point vertex is
proportional to $(p-p')^2$ (recall that $p_+-p_+'=p_--p_-'=p-p'$).

Having established the leading-order behavior of $\Lambda_\mu$, we can
now estimate the integral in Eq.~(\ref{nn1}). As we just showed, $\Lambda_\mu$
contributes a factor $(p-p')^2$. This factor cancels the gluon propagator.
Then the integral over
$p_0$ and $p_0'$ can be easily done, because of the respective poles in
the operators $D (p)$ and $D(p')$. The resulting expression contains six
three-momentum factors implying that the integral is $o(\alpha_s^8)$.
Therefore, the electroweak correction to the decay width is negligible.
Of course, no conclusion can be drawn regarding the exact value of the
electroweak
correction, because such a high order is beyond the scope of first-order
perturbation theory.

In conclusion, we have calculated electroweak corrections to the decay rate
of toponium in the dominant mode $t\to W^+ b$. Our results extend those of
Kummer and M\"odritsch~\cite{rf4}, who used the Coulomb gauge to obtain a
perturbative expansion of the Bethe-Salpeter equation. We employed a covariant
gauge and perturbed around the instantaneous solution to the Bethe-Salpeter
equation, in analogy with the abelian case of positronium. We found that the
electroweak corrections are suppressed by at least six powers of $\alpha_s$.
This takes us beyond the realm of first-order perturbation theory, therefore we
cannot calculate the precise value of the correction. It is
important to extend our results by including other electroweak effects, such
as a Higgs boson exchange. The development of an efficient systematic method
for such calculations involving bound states would be of great
interest, in view of the significance of the cross-section for
threshold t\=t production at future colliders~\cite{rf1}.

\vskip 2in

\noindent{\Large {\bf Acknowledgements}}

I wish to thank Bennie Ward for illuminating discussions.

\end{document}